\documentstyle[12pt]{article}

\newcommand{\be}{\begin{equation}}
\newcommand{\ee}{\end{equation}}
\newcommand{\ba}{\begin{eqnarray}}
\newcommand{\ea}{\end{eqnarray}}
\newcommand{\no}{\nonumber}

\newcommand{\la}{\langle}
\newcommand{\ra}{\rangle}

\newcommand{\hd}{\hskip2mm}
\renewcommand{\theequation}{\thesection.\arabic{equation}}

\begin{document}
\begin{flushright}
UT-882 \\
hep-th/0005008 \\
\end{flushright}

\bigskip

\begin{center}
{\Large \bf Five-Dimensional Gauge Theories and Local Mirror Symmetry}
\end{center}

\bigskip

\vskip15mm

\begin{center}

Tohru Eguchi 

\bigskip

{\it Department of Physics, Faculty of Science,

\medskip

University of Tokyo, 

\medskip

Tokyo 113, Japan}

\bigskip

\bigskip

and

\bigskip

\bigskip

Hiroaki Kanno

\medskip

{\it Department of Mathematics, Graduate School of Science,

\medskip

Hiroshima University 

\medskip

Higashi-Hiroshima 739, Japan}

\end{center}

\bigskip

\begin{abstract}
We study the dynamics of 5-dimensional gauge theory on $M_4\times S^1$ 
by compactifying type II/M theory on degenerate Calabi-Yau manifolds. 
We use the local mirror symmetry and 
shall show that the prepotential of the 5-dimensional $SU(2)$ 
gauge theory without matter is given exactly by that of the type II string
theory compactified on the local ${\bf F}_2$, 
i.e. Hirzebruch surface ${\bf F}_2$ 
lying inside a non-compact Calabi-Yau manifold. It is shown that our
result reproduces the Seiberg-Witten theory at the 4-dimensional limit
$R\rightarrow 0$ ($R$ denotes the radius of $S^1$) and also the result of the
uncompactified 5-dimensional theory at $R\rightarrow \infty$. 

We also discuss $SU(2)$ gauge theory with $1\le N_f\le 4$ matter in 
vector representations and show that they are described by the geometry of 
the local ${\bf F}_2$ blown up at $N_f$ points. 
\end{abstract}

\newpage

\section{Introduction}

Recent developments in non-perturbative string theory and M theory 
have led to new 
insights into the relation between low energy field theory 
and string theory: 
it has been argued in particular that non-perturbative dynamics takes place in 
low energy field theory and higher gauge 
symmetries emerge
when compactifying Calabi-Yau and $K_3$ manifolds 
degenerate and some of their homology cycles vanish.
For instance, 
when $K_3$ surface develops an $A$-$D$-$E$ 
singularity, there appears an enhanced 
$A$-$D$-$E$ gauge symmetry 
in 6-dimensions. Similarly, when a family of ${\bf P}^1$'s shrinks 
to zero size 
along a rational curve in 
Calabi-Yau threefold, we obtain a 4-dimensional 
$SU(2)$ gauge theory with ${\cal N}=2$ supersymmetry.

 In this article we would like to study the dynamics of 
SUSY gauge theories in 5-dimensions by compactifying
the M theory on degenerate Calabi-Yau manifolds.
We construct an effective action of an $SU(2)$ gauge theory on a 
5-dimensional space $M_4\times S^1$
where $M_4$ is the Minkowki space and $S^1$ is a circle of radius $R$. 
In the limit of $R\rightarrow 0$ this theory reproduces the standard 
${\cal N}=2$ SUSY gauge theory in 4-dimensions, 
i.e. Seiberg-Witten theory \cite{SW}. Thus our 5-dimensional 
model gives an M-theoretic generalization of Seiberg-Witten theory 
by incorporating Kaluza-Klein excitations. 
In the opposite limit $R\rightarrow \infty$ our model
reduces to the gauge theory in uncompactified 5-dimensions $M_5$. 
Characteristic features of this theory have been studied
using the brane-probe 
picture \cite{Sei} and also from the point of view of classical geometry of 
collapsing del Pezzo surfaces \cite{MS,DKV,IMS} and the behavior of
the low-energy effective gauge coupling has been determined exactly. 

In this paper we would like to propose an exact solution of the
5-dimensional theory on $M_4\times S^1$
which reproduces the known results at both limits $R\rightarrow 0$ and 
$R\rightarrow \infty$. As it turns out, our prepotential follows directly
from that
of the type II string theory compactified on singular Calabi-Yau manifolds
using the method of local mirror symmetry \cite{KKV,Kl,KMV,CKYZ}.
In the case of pure $SU(2)$ gauge theory without matter 
our result is obtained from the type II theory compactified on the local 
${\bf F}_2$, i.e. Hirzebruch surface ${\bf F}_2$ 
lying inside a Calabi-Yau threefold which
is the canonical bundle over ${\bf F}_2$. Similarly, $SU(2)$ gauge 
theories with $N_f$ matter 
in vector representations are also obtained from the type II theory 
compactified on the local 
${\bf F}_2$ blown up at $N_f$ points ($0 \le N_f\le 4$).
We will find that our model at $R=\infty$ has an infinite bare coupling 
constant and yields a non-trivial interacting field 
theory in the infra-red limit.

The local mirror symmetry is a method of 
mirror symmetry adapted in the case of non-compact Calabi-Yau manifolds.
Suppose, for instance, we are given a compact Calabi-Yau threefold which is 
an elliptic fibration over ${\bf F}_n$. One considers the limit of the size of 
the fiber $t_E$ going to $\infty$. Then the resulting non-compact manifold is 
modeled by the local ${\bf F}_n$, i.e. ${\bf F}_n$ inside a Calabi-Yau with 
the normal 
bundle being given by the canonical bundle of ${\bf F}_n$. 
The limit of $t_E \rightarrow \infty$ may also be considered as the limit 
of shrinking ${\bf F}_n$ with the size of the fiber kept fixed.
${\bf F}_n$ ($n=0,1,2$) and its various blow ups are the del Pezzo surfaces 
and 
these are in fact the type of manifolds which featured 
in the geometrical interpretation of 5-dimensional gauge theory 
\cite{MS,DKV,IMS}.

\section{$SU(2)$ Gauge Theory without Matter}

Let us start from the case of 5-dimensional gauge theory without matter.
We consider the local ${\bf F}_2$ model which is described by the toric data 
given in
the Appendix. Following the standard procedure \cite{KKV,Kl,KMV,CKYZ}
one obtains a curve in the B-model given by
\be
P=a_0x+a_1x^2+a_2\zeta+a_3+a_4{1 \over \zeta}=0.
\ee
Introducing a new variable $y=a_2\zeta-a_4/\zeta$ the curve is rewritten as
\be
y^2=(a_1x^2+a_0x+a_3)^2-4a_2a_4.
\ee
Complex moduli of the B-model are defined by
\be
z_F={a_1a_3 \over a_0^2}, \hskip5mm z_B={a_2a_4\over a_3^2}.
\ee
If we choose $a_1=a_3=1, a_0=s, 4a_2a_4=K^4$, we find
\be
y^2=(x^2+sx+1)^2-K^4, \hskip5mm z_F={1 \over s^2}, \hskip3mm z_B={K^4 \over 4}.
\label{Neka}
\ee
(\ref{Neka}) is in fact the curve proposed by Nekrasov \cite{Nek} for
the description of 5-dimensional gauge theory and its properties 
have been studied in 
\cite{KO} in detail. 

If we introduce a variable $U$ which is the analogue of $u$ of the  
Seiberg-Witten solution, the parameter $s$ is written as
\be
s=2R^2U
\ee
where $R$ is the radius of $S^1$.
In terms of $U$ and $R$ the curve reads as
\be
y^2=(x^2-R^4(U^2-{1 \over R^4}))^2-K^4.
\label{Nekb}
\ee
By comparing (\ref{Nekb}) with the Seiberg-Witten curve 
$y^2=(x^2-u)^2-\Lambda^4$,
we find the correspondence between the parameters $U$ and $u$
\be
u \Longleftrightarrow R^2(U^2-{1 \over R^4}).
\label{Nekc}\ee
As we see from (\ref{Nekc}), $U$ variable 
describes two copies of the $u$-plane.
Strong coupling region $u\approx 0$ maps to $U\approx \pm 1/R^2$ and
thus the two copies are separated by a distance of order $1/R^2$. 

In the brane-probe
interpretation of Seiberg-Witten solution \cite{Sen,BDS}, $u$-plane is
identified as a local region around one of the four fixed points 
(O-7 planes) in type I' theory compactification to 8-dimensions on $T^2$
($u=0$ is identified as the location of the O-7 plane). 
Then the curve (\ref{Nekb}) describes a theory which 
contains two of these O-7 planes. Since the fixed plane acts like a 
reflecting mirror, 
D3-brane probe will possess an infinite number of mirror images when 
inserted into a background of two orientifold planes. 
These mirror images are separated by
distances $n/R^2, \hskip2mm n\in {\bf Z}$ 
and open strings connecting 
them generate Kaluza-Klein modes of supersymmetric gauge fields.
Thus the curve (\ref{Nekb}) effectively 
describes a theory on a 5-dimensional manifold $M_4\times S^1$ with $R$ being
the radius of $S^1$.
In the limit of $R\rightarrow 0$ one of the $u$-planes moves off to 
$\infty$ and the model (\ref{Nekb}) is expected 
to reduce to the Seiberg-Witten theory.

Periods of the B-model of local ${\bf F}_2$ (\ref{Neka}) is determined by 
solving 
differential equations (Gelfand-Kapranov-Zelevinskij (GKZ) system) 
associated with the toric data (see Appendix). Differential operators 
are given by 
\ba
&&{\cal L}_1=z_F(4\theta_{z_F}^2+2\theta_{z_F})+\theta_{z_F}(2\theta_{z_B}
-\theta_{z_F}), \\
&&{\cal L}_2=z_B(2\theta_{z_B}-\theta_{z_F}+1)(2\theta_{z_B}-\theta_{z_F})
-\theta_{z_B}^2, 
\ea
where
\be
\theta_{z_F}\equiv z_F{\partial\over \partial z_F}, \qquad
\theta_{z_B}\equiv z_B{\partial\over \partial z_B}.
\ee
These operators have a regular singular point at $z_F=z_B=0$: they possess 
\lq\lq single-log" solutions $\omega_F,\hskip1mm \omega_B$ 
behaving as $\omega_F=\log z_F +\cdots$ and $\omega_B=\log z_B +\cdots$ 
at $z_F,z_B\approx 0$. There also exists 
a \lq\lq double-log" solution $\Omega$ behaving as $\Omega= (\log z_F)^2
+(\log z_F) (\log z_B) +\cdots$.
We identify the two single-log solutions 
as the K\"{a}hler parameters $t_F,t_B$
of the A-model: $t_F$ represents the size of the ${\bf P}^1$ fiber 
of ${\bf F}_2$ and $t_B$ the size of its base ${\bf P}^1$. $t_F$ is given by
\ba
&&\hskip-10mm -t_F\equiv \omega_F=\log z_F
+\left[\sum_{n\ge 1,m\ge 0}
{2(2n-1)!\over (n-2m)!n!m!^2}z_F^nz_B^m
-\sum_{m\ge 1}{(2m-1)!\over m!^2}z_B^m\right], \no \\
&&=-2\log\left({1 \over \sqrt{4z_F}}
+\sqrt{{1 \over 4z_F}-1}\right) \no \\
&&~~~~+\left[\sum_{n\ge 1,m\ge 1}
{2(2n-1)!\over (n-2m)!n!m!^2}z_F^nz_B^m
-\sum_{m\ge 1}{(2m-1)!\over m!^2}z_B^m\right]. 
\label{mirra} 
\ea
Similarly, $t_B$ is given by
\ba
&&\hskip-10mm -t_B\equiv \omega_B=-2\log\left({1 \over \sqrt{4z_B}}
+\sqrt{{1\over 4z_B}-1}\right).
\label{mirrb}\ea

In our interpretation as the 5-dimensional gauge theory,
the size of the fiber $t_F$ is identified as 
the vacuum expectation value $A$ of the scalar field in the vector-multiplet 
\be
t_F=4RA.
\label{mirrc}\ee
On the other hand, the size of the base 
$t_B$ is related to the dynamical mass 
parameter $\Lambda$ as
\be
e^{-t_B}=4R^4\Lambda^4 \hskip1mm .
\label{mirrd}\ee
Then the mirror transformation (\ref{mirrb}) becomes
\be
K={2R\Lambda\over \sqrt{1+4R^4\Lambda^4}} \hskip2mm .
\label{mirre}\ee
Note that (\ref{mirrc}) and (\ref{mirrd}) are in fact the identification 
of variables suggested in \cite{KKV,LN}.

One can invert the relations (\ref{mirra}),(\ref{mirrb}) perturbatively
and express the B-model parameters $z_F$,$z_B$ in terms of 
$t_F,t_B$. In the case of local mirror symmetry the
holomorphic solution of GKZ system is a constant and the mirror 
transformation is simpler than in the compact Calabi-Yau case.
One may then represent the double-log solution in terms of the K\"{a}hler
parameters
\ba
&&\hskip-5mm \Omega=t_F^2+t_Ft_B+4\sum_{n=1}{1 \over n^2}q_F^n
+4q_B\left(\sum_{n=1}n^2q_F^n\right) \no \\
&&~~\hskip2mm +q_B^2\big(q_F^2+36q_F^3+260q_F^4+1100q_F^5
+\cdots\big)+{\cal O}(q_B^3), \label{prepa} 
\ea
where
\be
q_F=e^{-t_F}, \qquad q_B=e^{-t_B}. 
\ee
We identify the double-log solution as $A_D$, the dual of the
variable $A$
\be
\Omega=-8\pi iRA_D.   
\label{prepe}
\ee

First two terms of (\ref{prepa}) represent the 
classical intersection numbers of the Calabi-Yau manifold and the 
remaining terms represent the contribution of world-sheet instantons.
According to ref \cite{CKYZ}
the double-log solution has a generic form
\be
\Omega=\sum_{i,j=1}^2t_it_j\la J^iJ^j\ra + \sum_{k=1}\sum_{n,m\ge 0}(\sum_i
x^i\partial_{t_i})d_{n,m}
{q_1^{kn}q_2^{km}\over k^3}~,
\label{prepb}\ee
for a local model of a surface $S$ (with two K\"{a}hler parameters).
$J_i$ denotes the K\"{a}hler classes of $S$ and $\la J_iJ_j\ra$ their 
intersection numbers. Numerical coefficients $x^i$ are defined by
\be
c_1(S)=\sum_ix^iJ_i~,
\ee
where $c_1(S)$ is the first Chern class of $S$. $d_{n,m}$ gives the number of 
rational holomorphic curves in the homology class $nJ_1+mJ_2$. Sum over $k$ in 
(\ref{prepb}) represents the multiple-cover factor.

By comparing (\ref{prepa}) and (\ref{prepb}) (set $q_1=q_F,q_2=q_B$) 
we find \cite{KKV,CKYZ}
\ba
&&d_{1,0}=-2, \hskip2mm d_{n,0}=0, \hskip1mm n>1, \hskip3mm d_{n,1}=-2n, \no \\
&&d_{1,2}=d_{2,2}=0,\hskip3mm d_{3,2}=-6, \hskip3mm d_{4,2}=-32, \cdots
\ea

By integrating $A_D$ over $A$ we have the prepotential $F$ for the local 
${\bf F}_2$ model
\ba
&&F={1\over 32\pi iR^2}\left[-{t_F^3\over 3}-{t_F^2t_B \over 2}
+4\sum_{n=1}{1 \over n^3}q_F^n
+4q_B\left(\sum_{n=1}nq_F^n\right)\right. \label{prepc} \no \\
&&\hskip5mm \left.+q_B^2\big({1 \over 2}q_F^2+12q_F^3+65q_F^4+220q_F^5
+\cdots\big)+{\cal O}(q_B^3)\right].
\ea

\section{Small and Large Radius Limits}
\setcounter{equation}{0}
Let us next examine the small and large radius 
limits $R \rightarrow 0,\infty$ of (\ref{prepc}). 
We first consider the 4-dimensional limit $R\rightarrow 0$. 
Due to the relations (\ref{mirrc}), (\ref{mirrd})
small $R$ corresponds to the base of ${\bf F}_2$ 
becoming large ($t_B\rightarrow \infty$) 
and the fiber becoming small ($t_F \rightarrow 0$) while the ratio
$e^{-t_B}/t_F^4$ is kept fixed. 
At small $R$, we have $K \approx 2R\Lambda$ and 
$U\approx \cosh(2RA)$. 
As explained by Katz, Klemm and Vafa \cite{KKV},
quantum parts of $F$ are suppressed because of the powers of
$q_B\approx R^4$ and the 
surviving contributions come from the divergent parts of
the series $\sum d_{n,m}q_F^n$ over $q_F$ as $q_F=e^{-4RA}\rightarrow 1$. 
At small $R$ the gauge coupling $\tau=\partial^2 F/\partial A^2$ behaves as
\ba
&&\tau={4i \over \pi}RA-{i \over 2\pi}\log(4R^4\Lambda^4)-{2i \over \pi}
\sum_{n=1}{q_F^n \over n}
-{8i \over \pi}R^4\Lambda^4\sum_{n=1}n^3q_F^n+\cdots \no \\
&&~~\approx {2i \over \pi}\log(2\sqrt{2}{A \over \Lambda})-{3i \over 16\pi}
{\Lambda^4 \over A^4}+\cdots
\label{prepsw}\ea
(\ref{prepsw}) reproduces the one-loop beta function and the one-instanton
contribution to the Seiberg-Witten solution \cite{SW}. 
We can check the agreement with
Seiberg-Witten theory for higher instanton terms. 

In (\ref{prepsw}) the world-sheet instanton 
expansion of type II theory is converted into
the space-time instanton expansion of gauge theory in the $R\rightarrow 0$ 
limit. Coefficients of the $m$-instanton amplitudes of gauge theory 
are determined by the asymptotic behavior of the number of holomorphic curves
$d_{n,m}$ as $n \rightarrow \infty$ with fixed $m$. 

Let us next examine the $R\rightarrow \infty$ limit of the uncompactified 
5-dimensional gauge theory. It is known \cite{Fer,Witten} that the gauge
theory on $M_5$ has no quantum corrections and its gauge coupling is
simply expressed in terms of classical intersection numbers of Calabi-Yau
manifold. In fact by taking $R\rightarrow \infty$ world-sheet
instanton terms disappear and we have
\be
\lim_{R\rightarrow \infty}{\tau \over 2\pi iR}\equiv \tau_5
={2 \over \pi^2}A~,
\label{Rinfa}\ee
(we have rescaled $\tau$ so that $\tau_5$ corresponds to the gauge coupling 
of 5-dimensional theory, $\tau_5=1/g^2_5$).
In the next section we will discuss local ${\bf F}_2$ model blown up at $N_f$ 
points. We then find that the above
formula is generalized as
\be
\tau_5={2 \over \pi^2}\left(A-\sum_{i=1}^{N_f}{1 \over 16}|A-M_i|
-\sum_{i=1}^{N_f}{1 \over 16}|A+M_i|\right).
\label{Rinfb}\ee
(\ref{Rinfb}) is exactly the behavior of 
gauge coupling of $SU(2)$ theory with $N_f$ matter in vector representations
(at infinite bare coupling) \cite{Sei,MS,DKV}. 
Thus we reproduce correct results also in the uncompactified
limit $R\rightarrow \infty$.

We should note that the local models of ${\bf F}_0$ and ${\bf F}_1$ 
also reproduce 
Seiberg-Witten solution in the $R\rightarrow 0$ limit \cite{KKV}
since the asymptotic 
behavior of the number of holomorphic curves $d_{n,m}$ are the same for all 
models ${\bf F}_i \hskip1mm , i=0,1,2$. ${\bf F}_0,{\bf F}_1$, 
however, have a different 
classical topology from ${\bf F}_2$ and do not reproduce (\ref{Rinfa}) at 
$R=\infty$. Thus the ${\bf F}_2$ model is singled out as 
the unique candidate for the description of 5-dimensional gauge theory
on $M_4\times S^1$.

\section{$SU(2)$ theory with matter}
\setcounter{equation}{0}

Let us next consider the case of $SU(2)$ gauge theory coupled to $N_f$ matter 
($1\le N_f\le 4$) in vector representations.
We first discuss the $N_f=1$ case. Relevant geometry is given by the local 
${\bf F}_2$ with one-point blown up. 
Corresponding curve is given by (see Appendix)
\be
y^2=(x^2+sx+1)^2-K^3(xw+{1 \over w}).
\ee
Comparison with the $N_f=1$ Seiberg-Witten curve suggests the identification
\be
w=e^{RM}~,
\ee
where $M$ denotes the bare mass of the matter multiplet.
Complex moduli of the B-model are given by (we denote 
$z_{F-E}$ as $z_F$ and $z_{F+E}$ as $z_{F'}$ for notational simplicity)
\be
z_F={w^2 \over s}, \hskip3mm z_{F'}={1 \over sw^2}, 
\hskip3mm z_B={K^3 \over 4w}.
\ee 
In this case the GKZ system is given by five partial differential equations 
in three variables (see Appendix) and 
is of considerable complexity. Here we content ourselves with the analysis of 
prepotential at the tree and one-loop level ignoring the space-time instantons.
This suffices for our purpose of extracting the $R\rightarrow \infty$ 
behavior of the theory. The small $R$ behavior has already been 
studied in \cite{Kl} and argued to reproduce Seiberg-Witten theory 
(we have also verified that the 1-instanton term is correctly
reproduced).

When we ignore instantons, single log solutions are given by
\ba
&&-t_F\equiv\omega_F=
\log(z_F)+\sum_{n=1}{(2n-1)!\over n!^2}(z_Fz_{F'})^n,\label{mirrnf1a}\\
&&-t_{F'}\equiv \omega_{F'}=
\log(z_{F'})+\sum_{n=1}{(2n-1)!\over n!^2}(z_{F}z_{F'})^n,\label{mirrnf1b}\\
&&-t_B\equiv \omega_B=\log(z_B).
\label{mirrnf1c}\ea
Identification with the variables of the gauge theory is given by
\be
t_F=2R(A-M), \hskip3mm t_{F'}=2R(A+M), \hskip3mm e^{-t_B}={2R^3\Lambda^3\over w}.
\ee
Then by inverting relations (\ref{mirrnf1a}),(\ref{mirrnf1b}) we find
\be
z_F={e^{-2R(A-M)}\over 1+e^{-4RA}},\hskip2mm z_{F'}={e^{-2R(A+M)}\over 
1+e^{-4RA}}.
\label{invertnf1}\ee
The double log solution 
\ba
&&\Omega=t_F^2+{1 \over 2}t_{F'}^2+2t_Ft_{F'}+t_B(t_F+t_{F'})
+\sum_{n=1}{4(2n-1)!\over n!^2}\sum_{j=1}^n{1 \over j}\hskip1mm 
(z_Fz_{F'})^n \no \\
&&\hskip10mm +\left(-\sum_{n>\ell}+\sum_{n<\ell }\right)
{(n+\ell-1)!\over n!\ell!}{1 \over n-\ell}
z_F^nz_{F'}^{\ell}~,
\ea
can then be re-expressed as
\ba
&&\hskip-5mm\Omega
=t_F^2+{1 \over 2}t_{F'}^2+2t_Ft_{F'}+t_B(t_F+t_{F'}) \no \\
&&~~~+4\sum_{n=1}{1\over n^2}e^{-4nRA}
-\sum_{n=1}{1 \over n^2}e^{-2nR(A+M)}-\sum_{n=1}{1 \over n^2}e^{-2nR(A-M)}~,
\label{adnf1}\ea
using (\ref{invertnf1}).

When we take the derivative in $A$, (\ref{adnf1}) gives
\be
\tau={i \over \pi}\left[2\log\sinh(2RA)
-{1 \over 4}\log\sinh(R(A+M))\sinh(R(A-M))\right].
\label{taunf1b}
\ee
In the 4-dimensional limit (\ref{taunf1b}) becomes
\be
\tau \approx{i \over \pi}(2-{1 \over 2})\log A
\ee
which gives the 1-loop beta function of Seiberg-Witten theory. 
On the other hand
in the 5-dimensional limit we find
\be
\tau_5=\lim_{R\rightarrow \infty}{\tau\over 2\pi iR}=
{2 \over \pi^2}\left[A-{1 \over 16}|A-M|-{1 \over 16}|A+M|\right]
\label{taunf1c}\ee
as we have claimed before. We note that in formulas 
(\ref{Rinfa}) and (\ref{taunf1c})
$\tau_5$ does not contain an additive constant which is identified as 
the bare coupling constant $\tau_{5,B}=1/g_{5,B}^2$. 
Thus the theory sits at the infinite bare coupling 
constant limit $g_{5,B}^2=\infty$ which yields non-trivial 5-dimensional 
theory in the infra-red regime \cite{Sei}.

We can similarly study the system with more matter up to $N_f=4$ using
the local mirror symmetry. Curves of the B-model are given by (see Appendix)
\be
y^2=(x^2+sx+1)^2-K^{4-N_f}\prod_{i=1}^{N_f}(w_ix+{1 \over w_i})~,
\label{curvematt}\ee
where parameters $w_i$ correspond to the bare masses of the matter multiplets
\be
w_i=e^{RM_i }, \hskip3mm i=1,\cdots,4
\ee
(\ref{curvematt}) agrees with the curve suggested by \cite{Nek} 
and \cite{BISTY}
(at $N_f=4$ the factor $K^{4-N_f}$ should be replaced by a dimensionless 
parameter $q$).  

In the cases $N_f\ge2$ the analysis of GKZ system becomes further 
involved: we will instead use a simpler method based on the 
Picard-Fuchs (PF) equation derived for the elliptic curves 
(\ref{curvematt}). It turns out that the PF equation of the elliptic
curve is an 
ordinary differential equation of third order
in the variable $A$ and can be studied relatively easily. This is the method 
used in ref \cite{KO}. This system, however, 
is not complete unlike that of GKZ case. The periods are 
determined only 
up to integration constants and one can not precisely
fix the mirror transformation. This ambiguity, however, affects only the
quantum part of the computation and one still obtains precise results for 
the prepotential at the tree and one-loop level. 

First we note that 
the quadratic curve $y^2=ax^4+4bx^3+6cx^2+4dx+e$ is transformed into 
the Weierstrass form
\be
y^2=4x^3-g_2x-g_3~,
\label{weier}\ee
by the relation
\be
g_2=ae-4bd+3c^2, \hskip3mm g_3=ace+2bcd-ad^2-b^2e-c^3.
\ee
We regard $g_2,g_3$ as functions of the parameter $s$.
Periods $\omega$ of the elliptic curve (\ref{weier}) obey the PF equation
\cite{KTS,KLRY}
\be
\frac{d^2 \omega}{ds^2} + c_1(s) \frac{d \omega}{ds} + c_0(s) \omega = 0~,
\ee
where
\ba
&&c_1 = - \frac{d}{ds} \log \left( \frac{3}{2\Delta} (2g_2\frac{dg_3}{ds}
-3\frac{dg_2}{ds}g_3) \right)~, \\
&&c_0 = \frac{1}{12} c_1 \frac{d}{ds} \log \Delta + \frac{1}{12}
\frac{d^2}{ds^2}
\log \Delta - \frac{1}{16 \Delta} \left( g_2 (\frac{dg_2}{ds})^2
-12 (\frac{dg_3}{ds})^2 \right)
\ea
and $\Delta = (g_2)^3 -27(g_3)^2$ denotes the discriminant of the curve.
Since $d A/d s$ is one of the periods of the curve, it satisfies the 
PF equation. Regarding $s$ as a function of $A$, we obtain 
\be
\frac{ds}{dA}\frac{d^3 s}{dA^3} - 3(\frac{d^2 s}{dA^2})^2 +
c_1 (\frac{ds}{dA})^2 \frac{d^2 s}{dA^2} - c_0 (\frac{ds}{dA} )^4 =0~.
\label{picarda}
\ee
This determines $s$ in terms of $A$.

Similarly $dA_D/ds=dA/ds \cdot d^2 F/d A^2$ satisfies the PF equation and
we obtain
\be
(\frac{ds}{dA})^{-1} \frac{d^4 F}{dA^4} -3  (\frac{ds}{dA})^{-2}
\frac{d^2 s}{dA^2}
\frac{d^3 F}{dA^3} + c_1\frac{d^3 F}{dA^3} =0~.
\ee
This equation can be integrated once and we find
\be
\frac{d^3 F}{dA^3} = \frac{const.}{\Delta} \left(2g_2\frac{dg_3}{ds}
-3\frac{dg_2}{ds}g_3\right) \left( \frac{ds}{dA}\right)^3.
\label{pfforf}\ee
Solving (\ref{pfforf}) determines the prepotential (const. in the 
right-hand-side is fixed by a suitable normalization of $F$).

In the case $N_f=2$, we find the gauge coupling at the classical and one-loop 
level as
\be
\tau={i \over \pi}\left[2\log\sinh(2RA)
-{1 \over 4}\prod_{i=1}^2\log\sinh(R(A+M_i))\sinh(R(A-M_i))\right].
\label{taunf2a}
\ee
In the 4-dimensional limit (\ref{taunf2a}) reads as 
\be
\tau\approx {2i \over \pi}(2-{1 \over 2}\cdot 2)\log A~,
\ee
which gives the beta function of $N_f=2$ Seiberg-Witten theory. 
On the other hand, in the five-dimensional limit we have 
\be
\tau_5={2 \over \pi^2}\left(A-\sum_{i=1}^2{1\over 16}|A-M_i|
-\sum_{i=1}^2{1 \over 16}|A+M_i|\right)
\ee
in agreement with (\ref{Rinfb}). 

We have checked that the local model of ${\bf F}_2$ blown up at 3 and 
4 points (see Appendix) also reproduce (\ref{Rinfb}). Thus we have obtained 
a model which appears to describe correctly 
the physics of 5-dimensional theory up to 
the number of flavors $N_f=4$ making use of the mirror symmetry.

\section{Discussions}

Since ${\bf F}_2$ is obtained from ${\bf F}_1$ which is 
a one-point blow up of ${\bf P}^2$, 
we effectively have up to 5-point blow ups of ${\bf P}^2$ describing 
the gauge theory. 
Unfortunately, beyond the 5-point blow up mirror symmetry of del Pezzo 
surfaces can not be described by toric geometry
and we can not apply our analysis for these cases.
In fact in the range $N_f\ge 5$ something drastic must happen, 
since in this asymptotically non-free region 
Seiberg-Witten solution does not exist and we should not have a smooth 
4-dimensional limit. On the other hand, in this range the 5-dimensional gauge
theories are expected to possess $E_n, \hskip1mm n=6,7,8$ global symmetries 
and are of particular interests. It is a challenging problem to clarify
the physics of gauge theory with $E_n$ symmetry. It may shed some light 
on the nature of asymptotically non-free field theories in 4-dimensions.

Our result shows that from the point of view of the type II/M theory 
compactified on Calabi-Yau manifold, low energy 5-dimensional theory on 
$M_4\times S^1$ emerges most
naturally with its prepotential being exactly the same as that of the 
string theory. On the other hand, 4-dimensional Seiberg-Witten theory 
appears only in the fine-tuned limit of the K\"{a}hler parameters. 
It seems possible 
that our 5-dimensional model is an example of an \lq\lq M-theoretic" 
lift of various 4-dimensional quantum field theories. We may imagine 
most of the 4-dimensional SUSY field theories in fact have a lift to 
5-dimensions where the quantum effects of the loops and instantons are 
replaced by purely geometrical effects of world-sheet instantons. 
It will be also 
quite interesting to see if there is a further lift of quantum field
theory to 6-dimensions as suggested by the duality between F- and M-theory.

\vskip1cm

We would like to thank M. Jinzenji, M. Naka and Y. Ohta
for discussions.
Researches of T.E. and H.K. are supported 
by the fund for the Special 
Priority Area No.707, Japan Ministry of Education.
We also acknowledge the stimulating atmosphere at Summer Institute 
'99 where this research was started. 

\vskip1cm
 
\section*{Appendix}
\renewcommand{\thesubsection}{A.\arabic{subsection}}
\renewcommand{\theequation}{A.\arabic{equation}}
\setcounter{equation}{0}

\subsection{ $N_f=0$ }

For pure Yang-Mills case we take the Hirzebruch surface ${\bf F}_2$ lying
inside a Calabi-Yau manifold.
Its toric diagram has five vertices;
\be
\nu_0 = (0,0)~, \quad \nu_1 = (1,0)~, \quad \nu_2 = (0,1)~, 
\quad \nu_3 = (-1,0)~, \quad \nu_4 = (-2,-1)~.
\ee
This is a two dimensional reflexive polyhedra no.4 in Fig.1 of \cite{CKYZ}.
The charge vectors satisfying the linear relation
\be
\sum_i \ell^{(k)}_i \nu_i = 0
\ee
are given by
\be
\ell^{(F)} = (-2 ; 1, 0, 1, 0 )~,  \quad
\ell^{(B)} = ( 0 ; 0, 1, -2, 1 )~.  
\ee
The constraint among B-model variables $Y_i, \hskip1mm i=0,1,\cdots,4$ 
\be
\prod_{\ell_i^{(k)}>0} Y_i^{\ell_i^{(k)}} = \prod_{\ell_i^{(k)}<0}
Y_i^{-\ell_i^{(k)}}
\ee
gives
\be
Y_0^2 = Y_1 Y_3~, \qquad Y_2 Y_4 = Y_3^2~,
\ee
and we have a solution
\be
Y = (sx, x^2, \zeta, s^2, \frac{s^4}{\zeta})~.
\ee
Set $s=1$. Then we obtain the following curve in the B model side
\ba
P_{{N_f=0}} &=& \sum a_i Y_i~, \no \\
 &=& a_0 x + a_1 x^2 + a_2 \zeta + a_3 + a_4 \frac{1}{\zeta}=0~.
\ea
Introducing a new variable $y = a_2 \zeta -(a_4/\zeta)$, we can rewrite the
curve as
\be
y^2 = (a_1 x^2 + a_0 x + a_3)^2 - 4 a_2 a_4~. \label{curve0}
\ee
For each charge vector $\ell^{(k)}$, a corresponding complex structure 
modulus is given by
\be
z_k = \prod_{i} a_i^{\ell^{(k)}_i}~.
\ee
In the present case we have
\be
z_F = \frac{a_1 a_3}{a_0^2}~, \qquad z_B = \frac{a_2 a_4}{a_3^2}~.
\ee
By setting
\be
a_1=a_3=1~, \quad a_0 = s~, \quad 4a_2 a_4 = K^4~,
\ee
we have
\be
z_F = \frac{1}{s^2}~, \qquad z_B = {K^4 \over 4}.
\ee
GKZ system is defined by a system of differential operators
\be
\prod_{\ell_i^{(k)}>0} 
\left({\partial \over \partial a_i}\right)^{\ell_i^{(k)}} 
= \prod_{\ell_i^{(k)}<0}
\left({\partial \over \partial a_i}\right)^{-\ell_i^{(k)}}~,
\ee
In the ${\bf F}_2$ case it is given by
\ba
&&{\cal L}_1=z_F(4\theta_{z_F}^2+2\theta_{z_F})+\theta_{z_F}(2\theta_{z_B}
-\theta_{z_F}), \\
&&{\cal L}_2=z_B(2\theta_{z_B}-\theta_{z_F}+1)(2\theta_{z_B}-\theta_{z_F})
-\theta_{z_B}^2, \\
&&~~\hskip3mm \theta_{z_i}\equiv z_i{\partial\over \partial z_i}, 
\hskip1mm i=F,B. \no
\ea

\subsection{ $N_f=1$ }

It is known that matters in the vector representation are generated by 
blowing up the manifold (see, for intance \cite{KV}). 
We expect that a blow up of the local ${\bf F}_2$ provides the
description of gauge theory with $N_f=1$ matter.
The corresponding 
reflexive polyhedra is no.6 of \cite{CKYZ} and given by the vertices
\ba
\nu_0 &=& (0,0)~, \quad \nu_1 = (1,0)~, \quad \nu_2 = (0,1)~, \no \\
\quad \nu_3 &=&(-1,1)~, \quad \nu_4 = (-1,0)~, \quad \nu_5 = (-1,-1)~.
\ea
We see that the charge vectors are
\ba
\ell^{(B)} = (0; 0, 0, 1, -2, 1)~, \no \\
\ell^{(F-E)} = (-1; 0, 1, -1, 1, 0)~, \no \\
\ell^{(E)} = (-1; 1, -1, 1, 0, 0)~. 
\ea
The constraint
\be
Y_3 Y_5 = Y_4^2~, \quad Y_0 Y_3 = Y_2 Y_4~, \quad Y_0 Y_2 = Y_1 Y_3~,
\ee
is solved by
\be
Y = (sx, sx^2, \frac{tx}{\zeta}, \frac{s}{\zeta}, s, \zeta s)~.
\ee
Setting $s=1$ gives the curve
\be
P_{N_f=1}= a_0 x + a_1 x^2 + a_2 \frac{x}{\zeta} + a_3 \frac{1}{\zeta}
+ a_4 + a_5 \zeta~,
\ee
or
\be
y^2 = (a_1 x^2 + a_0 x + a_4)^2 - 4 a_5 ( a_2 x + a_3)~.
\ee
If we set 
\be
a_1=a_4=1~, \quad a_0 = s~, \quad 4 a_5 = K^3~,
\quad a_2 = w~, \quad a_3 = w^{-1}~,
\ee
we obtain the curve \cite{Nek}
\be
y^2=(x^2+sx+1)^2- K^3(wx+{1 \over w})
\ee
Complex moduli are given by
\be
z_B= \frac{a_3 a_5}{a_4^2} = \frac{K^3}{4w}~,
\quad
z_{F-E} = \frac{a_2 a_4}{a_0 a_3} = \frac{w^2}{s}~,
\quad
z_{F+E} = \frac{a_1 a_3}{a_0 a_2} = \frac{1}{sw^2}~.
\ee
Complete set of differential equations is given by \cite{CKYZ} 
\ba
&&\hskip-18mm {\cal L}_1=\theta_B(\theta_B-\theta_{F-E}+\theta_{F+E})
- z_B(2\theta_B-\theta_{F-E})(2\theta_B-\theta_{F-E}+1), \\
&&\hskip-18mm {\cal L}_2=(\theta_{F-E}-\theta_{F+E})(\theta_{F-E}-2\theta_B)
-z_{F-E}(\theta_{F-E}+\theta_{F+E})(\theta_{F-E}-\theta_B-\theta_{F+E}), \\
&&\hskip-18mm {\cal L}_3=\theta_{F+E}(\theta_{F+E}+\theta_B-\theta_{F-E})
- z_{F+E}(\theta_{F+E}+\theta_{F-E})(\theta_{F+E}-\theta_{F-E}), \\
&&\hskip-18mm {\cal L}_4=(\theta_{F-E}-\theta_{F+E})\theta_B
- z_B z_{F-E}(\theta_{F-E}+\theta_{F+E})(2\theta_B-\theta_{F-E}), \\
&&\hskip-18mm {\cal L}_5=\theta_{F+E}(\theta_{F-E}-2\theta_B)
-z_{F-E} z_{F+E}(\theta_{F-E}+\theta_{F+E})(\theta_{F-E}+\theta_{F+E}+1).
\ea
We note that the last two operators are necessary to obtain
a unique double log solution $\Omega$.

\subsection{ $N_f=2$ }

We choose the following vertices
\ba
\nu_0 &=& (0,0)~, \quad \nu_1 = (1,0)~, \quad \nu_2 = (0,1)~, 
\quad \nu_3 =(-1,1)~, \no \\
\nu_4 &=& (-1,0)~, \quad \nu_5 = (-1,-1)~, 
\quad \nu_6 = (1,1)~.
\label{nf2poly}\ea
This is a reflexive polyhedra no.8 of \cite{CKYZ}.
The curve is given by
\be
P_{N_f=2} = a_0 x + a_1 x^2 + a_2 \frac{x}{\zeta} + a_3 \frac{1}{\zeta} 
+ a_4 + a_5 \zeta + a_6 \frac{x^2}{\zeta}=0~,
\ee
or
\be
y^2 = (a_1 x^2 + a_0 x + a_4 )^2 - 4 a_5 (a_6 x^2 + a_2 x + a_3)~.
\ee
Substituting the relation
\ba
a_1 &=& a_4 = 1~, \quad a_0 = s~, \quad 4a_5 = K^2~, \no \\
a_6 &=& w_1 w_2~, \quad a_2 = \left( \frac{w_2}{w_1} + \frac{w_1}{w_2} 
\right)~, \quad a_3 = (w_1 w_2)^{-1}~,
\ea
we obtain the complex structure moduli
\ba
z_B &=& \frac{a_3 a_5}{a_4^2}= \frac{K^2}{4 w_1 w_2}~,
\quad z_{F_1} = \frac{a_2 a_4}{a_0 a_3} = \frac{w_1^2 + w_2^2}{s}~, \no \\
z_{F_2} &=& \frac{a_1 a_3}{a_0 a_2}= \frac{1}{s(w_1^2 + w_2^2)}~,
\quad z_{F_3} = \frac{a_1 a_2}{a_0 a_6} = \frac{w_1^2 + w_2^2}{s w_1^2w_2^2}~.
\ea

\subsection{$N_f=3$}

Polyhedron for the 3-point blow up is obtained 
by adding a vertex $\nu_7=(0,-1)$
to (\ref{nf2poly}) (no.12 of \cite{CKYZ}). 
Elliptic curve is given by
\be
P_{Nf=3}=a_0x+a_1x^2+a_2{x \over \zeta}+a_3{1 \over \zeta}+a_4+a_5\zeta
+a_6{x^2 \over \zeta}+a_7x\zeta=0.
\ee
or
\be
y^2=(a_1x^2+a_0x+a_4)^2-4(a_6x^2+a_2x+a_3)(a_7x+a_5).
\ee
By choosing the variables as
\ba
&&a_0=s,\hd a_1=1,\hd a_2=\frac{K}{4}({w_1 \over w_2}+{w_2 \over w_1}),
\hd a_3={K \over 4 w_1w_2}, \hd a_4=1,\hd  
a_5={1\over w_3}, \no \\
&&a_6=\frac{K w_1w_2}{4}, \hd a_7=w_3, 
\ea
we find the complex moduli
\ba
&&z_B={a_3a_5 \over a_4^2}={K \over 4 w_1w_2w_3},
\hskip3mm z_{F_1}={a_2a_4\over a_0a_3}={1\over s}
(w_1^2+w_2^2),
\hskip3mm z_{F_2}={a_1a_3\over a_0a_2}={1\over s}{1\over w_1^2+w_2^2}, \no \\
&&z_{F_3}={a_1a_2\over a_0a_6}
={1\over s}({1\over w_1^2}+{1\over w_2^2}),
\hskip3mm z_{F_4}={a_4a_7\over a_0a_5}={w_3^2\over s}.
\no \\
&&\ea

\subsection{$N_f=4$}

Polyhedron for the 4-point blow up
is obtained by further adding a vertex $\nu_8=(1,-1)$ to the polyhedron
of $N_f=3$ (no.15 of \cite{CKYZ}).
Elliptic curve is given by
\be
P_{N_f=4}=a_0x+a_1x^2+a_2{x \over \zeta}+a_3{1 \over \zeta}+a_4+a_5\zeta
+a_6{x^2 \over \zeta}+a_7x\zeta
+a_8x^2\zeta=0.
\ee
By eliminating $\zeta$ it becomes
\be
y^2=(a_1x^2+a_0x+a_4)^2-4(a_6x^2+a_2x+a_3)(a_8x^2+a_7x+a_5)~.
\ee
By choosing the variables as
\ba
&&a_0=s,\hd a_1=1,\hd a_2=({w_1 \over w_2}+{w_2 \over w_1})q,
\hd a_3={1 \over w_1w_2}q, \hd a_4=1,\hd  a_5={1\over w_3w_4}, \no \\
&&a_6=w_1w_2q, \hd a_7=({w_3\over w_4}+{w_4\over w_3}), \hd
a_8=w_3w_4 
\ea
we find the complex moduli
\ba
&&z_B={a_3a_5 \over a_4^2}={q \over w_1w_2w_3w_4},
\hskip3mm z_{F_1}={a_2a_4\over a_0a_3}={1\over s}
(w_1^2+w_2^2),
\hskip3mm z_{F_2}={a_1a_3\over a_0a_2}={1\over s}{1\over w_1^2+w_2^2}, \no \\
&&z_{F_3}={a_1a_2\over a_0a_6}
={1\over s}({1\over w_1^2}+{1\over w_2^2}),
\hskip3mm z_{F_4}={a_4a_7\over a_0a_5}={w_3^2+w_4^2\over s},\hskip3mm
z_{F_5}={a_1a_7\over a_0a_8}={1 \over s}({1\over w_3^2}+{1\over w_4^2}).
\no \\
&&\ea

\end{document}